\documentclass{article}
\usepackage{times}
\usepackage{amsfonts}
\usepackage{graphicx}
\begin{document}
\noindent
{\Large  ON THE COSMOLOGICAL CONSTANT AS A\\ QUANTUM OPERATOR}
\noindent

\vskip.5cm
\noindent
{\bf P. Fern\'andez de C\'ordoba}$^{a,1}$,  {\bf R. Gallego Torrom\'e}$^{b,2}$,
{\bf S. Gavasso}$^{a,3}$ and {\bf  J.M. Isidro}$^{a,4}$\\
$^{a}$Instituto Universitario de Matem\'atica Pura y Aplicada,\\ Universitat Polit\`ecnica de Val\`encia, Valencia 46022, Spain\\
$^{b}$Department of Mathematics, Faculty of Mathematics, Natural Sciences\\ and Information Technologies, University of Primorska, Koper, 
Slovenia\\
$^{1}${\tt pfernandez@mat.upv.es}, $^{2}${\tt rigato39@gmail.com},\\ 
$^{3}${\tt sgavas@alumni.upv.es}, $^{4}${\tt joissan@mat.upv.es}\\
\vskip.5cm
\noindent
\vskip.5cm
\noindent
{\bf Abstract}  We regard the cosmological fluid within an exponentially expanding FLRW spacetime as the probability fluid of a nonrelativistic Schroedinger field. The scalar Schroedinger particle so described has a mass equal to the total (baryonic plus dark) matter content of the Universe. This procedure allows a description of the cosmological fluid by means of the operator formalism of nonrelativistic quantum theory. Under the assumption of radial symmetry, a quantum operator proportional to $1/r^2$ represents the cosmological constant $\Lambda$. The experimentally measured value of $\Lambda$ is one of the eigenvalues of $1/r^2$. Next we solve the Poisson equation $\nabla^2U=\Lambda$ for the gravitational potential $U(r)$, with the cosmological constant $\Lambda(r)=1/r^2$ playing the role of a source term. It turns out that $U(r)$ includes, besides the standard Newtonian potential $1/r$, a correction term proportional to $\ln r$ identical to that appearing in theories of modified Newtonian dynamics.


\section{Introduction}\label{einfuehrung}

Understanding the physics of the dark Universe is a major challenge of current research. It is comparable in relevance to an older outstanding challenge, namely the search for a quantum theory of gravity. Given the deep interconnections between these two problems, one hopes that shedding light on either one of the two will also help to clarify the other. 

Even a tiny value of the cosmological constant  $\Lambda$ casts a long shadow, as explained in ref. \cite{ASHTEKAR} in the context of gravitational waves. For example, a nonvanishing $\Lambda$ prevents Minkowski space from being a vacuum solution to the Einstein field equations. Beyond its key role in the $\Lambda$CDM model \cite{WEINBERG2}, $\Lambda$ has adopted a large number of different impersonations. To name just a few: $\Lambda$ as the eigenvalue of a Sturm--Liouville problem \cite{RUSOS}; $\Lambda$ as a quantum field \cite{BARROW, ELIZALDE};  $\Lambda$ as a gravitational dipole \cite{BLANCHET}; $\Lambda$ as a parameter measuring deviation from some variant of classicality \cite{PAGANINI}; $\Lambda$ as a fluid \cite{CADONI}; $\Lambda$ as a close cousin of the quantum potential \cite{MATONE2}; $\Lambda$ as a Bose--Einstein condensate \cite{ITALIA}; $\Lambda$ as an event horizon \cite{GAZTANAGA}; $\Lambda$ as a free thermodynamical variable \cite{MAKELA};  $\Lambda$ as a Lagrange multiplier in the gravitational action \cite{SORKIN}; $\Lambda$ as the vacuum energy of particle physics \cite{WEINBERG1}; finally there are also intriguing proposals in connection with the modern approach of emergent gravity \cite{JACOBSON, PADDY1, SINGH2, VERLINDE1, VERLINDE2}. The previous list is necessarily incomplete; for further literature and detailed analysis we refer to the excellent reviews \cite{CARROLL, MARTIN, PEEBLES, SOLA}.

In upcoming work \cite{FINSTER2} a geometric model for the dark Universe within the framework of causal fermion systems \cite{FINSTER1} will be put forward. 
In this letter we propose yet another impersonation for the cosmological constant, namely a (nonrelativistic) quantum operator, one of whose eigenvalues will equal the experimentally measured value of $\Lambda$. To this end we must first explain the appearance of a (nonrelativistic) Hilbert space of quantum states, and how the cosmological constant arises naturally as an operator.

\section{The cosmological constant as a source term}

Our current Universe is fairly accurately described by a flat, Friedmann--Lema\^{\i}tre--Robertson--Walker geometry, with a small but positive cosmological constant $\Lambda$ \cite{WEINBERG2}. The cosmological fluid within this spacetime can be modelled as a perfect, Newtonian fluid. As such it is governed by the continuity equation and the Euler equation:
\begin{equation}
\frac{\partial\rho}{\partial t}+\nabla\cdot\left(\rho{\bf v}\right)=0,\qquad \frac{\partial{\bf v}}{\partial t}+\left({\bf v}\cdot\nabla\right){\bf v}+\frac{1}{\rho}\nabla p-{\bf F}=0.
\label{uno}
\end{equation}
Above, $\rho$ is the volume density of fluid matter, $p$ its pressure,  ${\bf v}$ the velocity field and ${\bf F}$ the external force acting per unit mass of the fluid. To the density $\rho$ there contribute the baryonic ($\sim 5\%$) and the dark ($\sim 27\%$) matter contents of the Universe. The remaining $\sim 68\%$ of the total mass/energy budget of the Universe is currently attributed to the dark energy represented by the cosmological constant $\Lambda$.

The cosmological fluid can be equivalently modelled as the probability fluid of a scalar Schroedinger particle. The latter is described by a wavefunction $\Psi$ that we decompose \` a la Madelung \cite{MADELUNG} into amplitude and phase as
\begin{equation}
\Psi=\exp\left(S+\frac{{\rm i}}{\hbar}I\right).
\label{dos}
\end{equation}
The phase $\exp({\rm i}I/\hbar)$ is the complex exponential of the classical--mechanical action integral $I$, and we have written the amplitude as the exponential of the real, dimensionless quantity $S$. It turns out that the following bijection can be established \cite{NOSALTRES1, NOSALTRES2, NOIFLRW, NOSALTRES3} between the cosmological fluid of Eq. (\ref{uno}) and the quantum mechanics of Eq. (\ref{dos}): 
\begin{equation}
\begin{tabular}{| c | c | c |}\hline
& Euler &  Madelung  \\ \hline
volume density& $\rho$ & $\exp(2S)$   \\ \hline
velocity & ${\bf v}$ & $\nabla I/M$   \\ \hline
pressure term & $\nabla p/\rho$ & $\nabla Q/M$   \\ \hline
external forces & ${\bf F}$ & $-\nabla V/M$   \\ \hline
\end{tabular}
\label{tabla}
\end{equation}
Above, $V$ is the potential energy entering the Schroedinger equation satisfied by $\Psi$, $M$ stands for the sum of the baryonic and dark matter contents of the Universe, and $Q$ is the quantum potential defined by
\begin{equation}
Q=-\frac{\hbar^2}{2M}\left[\left(\nabla S\right)^2+\nabla^2 S\right].
\label{quantistico}
\end{equation}
Using the dictionary (\ref{tabla}) we conclude that, given a Newtonian cosmological fluid, one can conveniently describe it as a nonrelativistic quantum mechanics.

It is known that an arbitrary vector field ${\bf F}$ is determined (up to additive constants) by its curl $\nabla\times{\bf F}$ and its divergence $\nabla\cdot{\bf F}$. For the external forces acting on the cosmological fluid we have ${\bf F}=-\nabla V/M$, hence $\nabla\times{\bf F}=0$. However the divergence $\nabla\cdot{\bf F}=-\nabla^2 V/M$ need not vanish. Setting $U=V/M$, we will be interested in the Poisson equation
\begin{equation}
\nabla^2U=\Lambda,
\label{puason}
\end{equation}
where we have posited 
\begin{equation}
\Lambda=-\nabla\cdot{\bf F}.
\label{definiamo}
\end{equation}

\section{The gravitational potential in FLRW space}

Starting from a knowledge of $U$, Eq. (\ref{puason}) defines the operator representing the cosmological constant $\Lambda$. Conversely, in this letter we will use the specific expressions for the operator $\Lambda$ obtained in refs. \cite{NOSALTRES1, NOSALTRES2, NOIFLRW, NOSALTRES3} under the assumption of radial symmetry, and integrate Eq. (\ref{puason}) to obtain $U$.

The analysis performed in refs. \cite{NOSALTRES1, NOSALTRES2, NOSALTRES3} applied to the Newtonian spacetimes $\mathbb{R}\times\mathbb{R}^3$, $\mathbb{R}\times\mathbb{H}^3$ and $\mathbb{R}\times\mathbb{S}^3$ respectively. However, the expressions therein obtained for $\Lambda$ easily carry over to the relativistic cases of flat, hyperbolic and spherical FLRW spacetimes, respectively.

\subsection{Flat space}

In flat FLRW spacetime we have the metric \cite{WEINBERG2}
\begin{equation}
{\rm d}s^2={\rm d}t^2-a^2(t)\left({\rm d}r^2+r^2{\rm d}\Omega^2\right), \qquad {\rm d}\Omega^2={\rm d}\theta^2+\sin^2\theta\,{\rm d}\phi^2.
\label{flachemetrik}
\end{equation}
We pick the exponentially expanding scale factor
\begin{equation}
a(t)=\exp[H_0(t-t_0)], 
\label{jabel}
\end{equation}
with $H_0$ the Hubble constant, as befits the current expansion of the Universe; in the present era we can set $t=t_0$. Under these circumstances we have in ref. \cite{NOSALTRES1} considered the quantum operator 
\begin{equation}
\Lambda=\frac{C_0}{r^2},
\label{landaplano}
\end{equation}
with $C_0$ a dimensionless constant. Substitution of Eq. (\ref{landaplano}) into (\ref{puason}) and integration yields
\begin{equation}
U_0(r)=C_0\ln \frac{r}{r_1}+\frac{r_2}{r},
\label{inteplano}
\end{equation}
with $r_1$, $r_2$ integration constants.

We see that, besides the standard Newtonian potential $1/r$, a logarithmic correction $\ln r$ arises.  Logarithmic corrections to the Newtonian potential have been reported in MOND gravity \cite{MILGROM} as an explanation for the flatness of galaxy rotation curves.

\subsection{Spherical space}

In spherical FLRW spacetime we have the metric \cite{WEINBERG2}
\begin{equation}
{\rm d}s^2={\rm d}t^2-R_0^2\,a^2(t)\left[{\rm d}\chi^2+\sin^2\chi\left({\rm d}\theta^2+\sin^2\theta{\rm d}\varphi^2\right)\right],
\label{catorce}
\end{equation}
where $R_0$ is the radius of the sphere, and the scale factor is chosen as in Eq. (\ref{jabel}). A radial coordinate $r$ is defined in terms of the angular coordinate $\chi$ as
\begin{equation}
r=R_0\sin\chi, \qquad 0<\chi<\pi, 
\label{cincuenta}
\end{equation}
and the operator representing the cosmological constant becomes \cite{NOSALTRES3}
\begin{equation}
\Lambda=\frac{C_+}{R_0^2\sin^2\chi},
\label{landaesfe}
\end{equation}
with $C_+$ a dimensionless constant. Substituting Eq. (\ref{landaesfe}) into (\ref{puason}) and integrating produces
\begin{equation}
U_+(\chi)=C_+\left(\ln\sin\chi-\chi\cot\chi\right)+\chi_1^+\cot\chi+\chi_2^+,
\label{inteesfe}
\end{equation}
with $\chi_1^+$, $\chi_2^+$ integration constants.

Again we observe the presence of a logarithmic correction to the radially symmetric Newtonian potential $\cot\chi$. Now the limit $R_0\to\infty$ should yield back flat FLRW space. Taylor--expanding in powers of $\chi$ and keeping only first--order terms  correctly reduces the spherical potential (\ref{inteesfe})  to the flat potential (\ref{inteplano}).

\subsection{Hyperbolic space}

Negatively--curved FLRW spacetime carries the metric \cite{WEINBERG2}
\begin{equation}
{\rm d}s^2={\rm d}t^2-R_0^2\,a^2(t)\left[{\rm d}\chi^2+\sinh^2\chi\left({\rm d}\theta^2+\sin^2\theta\,{\rm d}\varphi^2\right)\right],
\label{veinte}
\end{equation}
where $R_0$ is the radius of the hyperboloid, and the scale factor is again given by Eq. (\ref{jabel}). As before we can define a radial coordinate $r$ in terms of the angular coordinate $\chi$:
\begin{equation}
r=R_0\sinh\chi, \qquad 0<\chi<\infty.
\label{cientouno}
\end{equation}
The cosmological constant is represented by the operator \cite{NOSALTRES2}
\begin{equation}
\Lambda=\frac{{C}_-}{R_0^2}\frac{1}{\sinh^2\chi},
\label{sesenta}
\end{equation}
where $C_-$ is a dimensionless constant. As usual we susbtitute Eq. (\ref{sesenta}) into the right--hand side of Eq. (\ref{puason}) and integrate. We arrive at
\begin{equation}
U_-(\chi)=C_-\left(\ln\sinh\chi-\chi\coth\chi\right)+\chi_1^{-}\coth\chi+\chi_2^{-}
\label{intehyp}
\end{equation}
with $\chi_1^-$, $\chi_2^-$ integration constants.

Also in this case there is a logarithmic correction to the hyperbolic Newtonian potential $\coth\chi$. Furthermore, Taylor--expanding in powers of $\chi$ and keeping only first--order terms reduces the hyperbolic potential (\ref{intehyp}) to the flat potential (\ref{inteplano}).

\section{Discussion}

In this letter we have considered the usual three homogeneous and isotropic cosmological spacetimes: flat FLRW space, hyperbolic FLRW space, spherical FLRW space \cite{WEINBERG2}, all three of them assumed in exponential expansion as in the current Universe. The cosmological fluid has been modelled as the probability fluid of a nonrelativistic Schroedinger particle. Assuming the latter described by a wavefunction $\Psi$, the bijection between its probability fluid and the cosmological fluid is explained in table (\ref{tabla}). 

We have borrowed the expression for the cosmological constant as a quantum operator $\Lambda(r)$ derived in refs. \cite{NOSALTRES1, NOSALTRES2, NOIFLRW, NOSALTRES3}. This operator acts on the Hilbert space of quantum states $\Psi$. Finally we have set $\Lambda(r)$ as the source term in the Poisson equation $\nabla^2U(r)=\Lambda(r)$ and solved for $U(r)$, in each one of the three FLRW spacetimes considered. Above, $r$ stands for a radial coordinate, $\nabla^2$ is the radial piece of the Laplacian operator (with respect to the corresponding FLRW  metric), and $U(r)$ is the radially symmetric gravitational potential.

The resulting potentials are collected in Eqs. (\ref{inteplano}), (\ref{inteesfe}) and (\ref{intehyp}). All three solutions conform to the following pattern. First, they all contain a leading term proportional to the corresponding Newtonian potential.  Second, they all exhibit a subleading term, a logarithmic correction in the form of a summand proportional to $\ln r$. This correction term vanishes whenever the cosmological constant is set to zero. Moreover, this logarithmic correction is coincident with the corresponding term in MOND gravity \cite{MILGROM}.

We conclude that the ansatz for the dark energy made in Eqs. (\ref{landaplano}), (\ref{landaesfe}) and (\ref{sesenta}) naturally leads to the correct Newtonian potential for the dark matter of the Universe. This provides a strong consistency check of our proposal.

\end{document}